\documentclass[prl,aps,twocolumn,floatfix,showpacs]{revtex4}
\usepackage{graphics}
\usepackage{color}
\begin{document}
\title{Spectral densities of superconducting qubits with environmental resonances.}
\author{Kaushik Mitra, C. J. Lobb, and C. A. R. S{\'a} de Melo}
\affiliation{Joint Quantum Institute and Department of Physics \\ 
University of Maryland College Park MD 20742}
\date{\today}

\begin{abstract}

\end{abstract}
\maketitle

\begin{figure}[htb]
\centerline{
\scalebox{0.40}{
\includegraphics{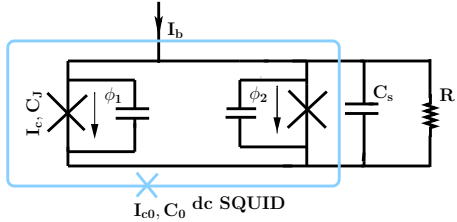}
}
}
\caption{
Flux qubit measured by a dc-SQUID (gray line). The qubit is in the inner 
SQUID loop with critical current $I_c$ and capacitance $C_J$ 
for both junctions.
The inner SQUID is shunted by a capacitance $C_s$, and environmental resistance $R$ 
and is biased by a ramping current $I_b$. The dc-SQUID loop has 
junction capatitance $C_0$ and critical current $I_{c0}$.}
\label{fig:one}
\end{figure}

\begin{figure}[htb]
\centerline{
\scalebox{0.30}{
\includegraphics{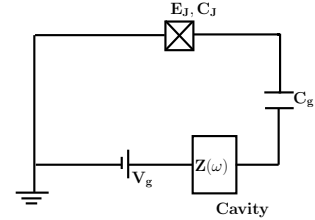}
}
}
\caption{Circuit diagram of the Cooper-pair box. The superconducting island 
is connected to a large reservoir through a 
Josephson junction with Josephson energy $E_J$ and capacitance $C_g$. 
The voltage bias $V_g$ is provided through a resonator (cavity) having
environmental impedance $Z(\omega)$, which is connected 
to $C_g$ as shown.
}
\label{fig:two}
\end{figure}

\begin{figure}
\centerline{
\scalebox{0.42}{
\includegraphics{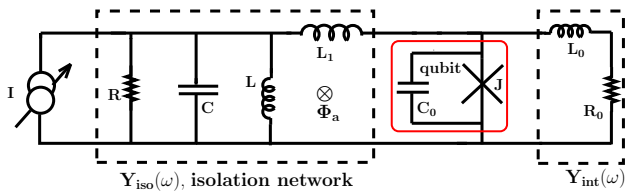}
}
}
\caption{Schematic drawing of the phase qubit with an RLC isolation circuit.}
\label{fig:three}
\end{figure}

In this paper we derive the environmental spectral density for
flux~\cite{ref:mooij}, phase~\cite{ref:martinis} and charge~\cite{ref:blais} qubits when each of them is 
coupled to an environment with a resonance. From the spectral density we obtain the characteristic 
spontaneous emission (relaxation) lifetimes $T_1$ for each of these qubits, and show that
there is a substantial enhancement of $T_1$ beyond the resonant frequency of the environment.
The circuits considered are shown in Fig.~\ref{fig:one},~\ref{fig:two} and ~\ref{fig:three}. 

The flux qubit shown in Fig.~\ref{fig:one} is measured by a 
dc-SQUID. Hence to study decoherence and relaxation time scales, one has 
to consider the noise that is transferred from the qubit to the dc-SQUID. 

The classical equation of motion for the dc-SQUID is 
\begin{equation}
\label{eqn:ceq1}
C_0 \ddot \phi   + \frac{2\pi}{\Phi_0}I_{c0}\sin\phi -\frac{2\pi}{\Phi_0}I + 
\int_0^t dt'  Y(t - t') \dot\phi (t') = 0
\end{equation}
where $\phi$ is the gauge invariant phase across the Josephson junction of the outer dc-SQUID loop, 
$I_{c0}$ is the critical current of its junction, $\Phi_0 = h/2e$ is the flux quantum. 
and the total induced current in the outer dc-SQUID is  
\begin{equation}
\label{eqn:I}
I = \frac{4}{L_{dc}} \langle\delta \phi_0 \sigma_z \rangle + \frac{4}{L_{dc}}\langle\phi_m\rangle +
\left(\frac{2\pi}{\Phi_0}\right)^2 J_1 \langle\phi_p\rangle.
\end{equation}
Here $\phi_p$ and $\phi_m$ are the sum and difference of the
gauge invariant phases across the junctions of the inner SQUID, 
$L_{dc}$ is the self-inductance of the inner SQUID,
and $J_1$ is the bilinear coupling between $\phi_m$ and $\phi_p$ at the potential energy minimum.
The term $\delta\phi_0 = \pi M_q I_{cir} / \Phi_0$, where $I_{cir}$ is the circulating current 
of the localized states of the qubit (described in terms of Pauli matrix $\sigma_z$), 
and $M_q$ is the mutual inductance between the qubit and the outer dc-SQUID.
The last term in Eq.~(\ref{eqn:ceq1}) is the dissipation term due to effective admittance
$Y(\omega)$ felt by the outer dc-SQUID. 

For the charge qubit shown in Fig.~\ref{fig:two} the classical equation of motion 
for the charge Q is 
\begin{eqnarray}
\label{eqn:ceq2}
V_g(\omega) =\left(-\frac{\omega^2 L_J(\omega)}{1-\omega^2 L_J(\omega) C_J} + \frac{1}{C_g} 
+ i\omega Z(\omega)\right)Q(\omega)
\end{eqnarray}
Here,  $V_g$ is the gate voltage, $C_g$ is the gate capacitance, $L_J$ and $C_J$ 
are the Josephson inductance and capacitance respectively, $Z(\omega)$ is the effective impedance 
seen by the charge qubit due to a transmission line resonator (cavity), 
$\Omega$ is the resonant frequency of the resonator, and $Q$ is the charge across $C_g$.  

For the phase qubit shown in Fig.~\ref{fig:three} the  classical equation of motion is
\begin{equation}
\label{eqn:ceq3}
C_0 \ddot \gamma   + \frac{2\pi}{\Phi_0}I_{c0}\sin\gamma -\frac{2\pi}{\Phi_0}I + 
\int_0^t dt'  Y(t - t') \dot\gamma (t') = 0
\end{equation}
where $I_{c0}$ is the critical current of Josephson junction $J$ in Fig.\ref{fig:one}, $I$ is 
the bias current, and $\Phi_0 = h/2e$ is the flux quantum.
The last term is the dissipation term due to $Y(\omega)$ 
which is the effective admittance as seen by the dc-SQUID. 

The equations of motion described in Eqs.~(\ref{eqn:ceq1}), (\ref{eqn:ceq2}), and (\ref{eqn:ceq3}),  
can be all approximatelly described by the effective spin-boson Hamiltonian 
\begin{equation}
\label{eqn:hamiltonian}
\widetilde{H} = \frac{\hbar \omega_{01}}{2}\sigma_z
+ \sum_k \hbar \omega_k b_k^\dagger b _k
+ H_{SB},
\end{equation}
written in terms of Pauli matrices $\sigma_i$ (with $i = x, y, z$) and boson 
operators $b_{k}$ and $b_{k}^\dagger$.
The first term in Eq.~(\ref{eqn:hamiltonian}) represents a two-level approximation
for the qubit (system) described by states $\vert 0 \rangle$ and $\vert 1 \rangle$ 
with energy difference $\hbar \omega_{01}$.
The second term corresponds to the isolation
network (bath) represented by a bath of bosons, where 
$b_{k}$ and $b_{k}^\dagger$ are the annihilation and creation operator of 
the $k$-th bath mode with frequency $\omega_{k}$. 
The third term is the system-bath (SB) Hamiltonian which corresponds to the coupling 
between the environment and the qubit. 

At the charge degeneracy point for the charge qubit (gate charge $N_g=0$),
at the flux degeneracy point (external flux $\Phi_{\rm ext}=\pi\Phi_0$) for the flux qubit, and
the suitable flux bias condition for the phase qubit (external flux $\Phi_{\rm a}=L_1\phi_0$)
$H_{SB}$ reduces to,
\begin{equation}
\label{eqn:system-bath-hamiltonian}
H_{SB} = 
\frac{1}{2}\sigma_x\hbar\langle 1\vert v\vert 0\rangle 
\sum_k \lambda_{k1} \left( b_{k}^\dagger +  b_{k} \right)
\end{equation}
where $v=\phi$ for the flux qubit, $Q$ for the charge qubit, and $\gamma$ for 
the phase qubit. The spectral density of the bath modes $J(\omega) = 
\hbar \sum_k \lambda_k^2  \delta \left( \omega- \omega_{k} \right)$ has dimensions of energy and can be written
as $J(\omega) = \omega {\rm Re} Y (\omega) (\Phi_0/2\pi)^2$ for flux and phase qubits and  
$J(\omega) = 2 \hbar\omega e^2/\hbar{\rm Re} Z (\omega)$ for charge qubits. 

%
%

%

For the flux qubit circuit shown in Fig.~\ref{fig:one}, 
the shunt capacitance $C_s$ is used to control the environment, 
while the Ohmic resistance of the circuit is modelled by $R$. 
In this case, the environmental spectral density is 
\begin{equation}
\label{eqn:spectral-density}
J_{1}\left(\omega\right) =
\frac{\alpha_1\omega}{\left(1-\omega^2/\Omega_1^2\right)^2+4\omega^2\Gamma_1^2/\Omega_1^4}.
\end{equation}
when $\omega_m \gg {\rm max}(\omega_p, \omega_0)$, and 
%
%
when the dc-SQUID is far away from the switching point 
to be modelled by an ideal inductance $L_J$. 
Here, 
$\Omega_1 = 1/\sqrt{L_J C_s} = \sqrt{ 4 \pi I_{c} /(C_s \Phi_0) }$ 
is the plasma frequency of the inner SQUID and plays the role of
the resonant frequency, 
where $I_{c}$ is the critical current for each of two Josephson junctions. 
Also, $\Gamma_1=  {1}/(C_s R)$ plays the role of the resonance width, and
$\alpha_1  =  2 (e I_p I_b M_q)^2 /(C_s^2 \hbar^2 R \Omega_1^4)$ reflects
the low frequency behavior. 
The coupling between the flux qubit and the outer dc-SQUID occurs emerges from
the interaction of the persistent current $I_p$ of the qubit 
and the bias current $I_b$ of the dc-SQUID via their mutual inductance $M_q$.

The spectral density for the charge qubit shown in Fig.~\ref{fig:two}
is obtained by solving for the normal modes of the resonator and transmission lines, 
including an input impedance
$R$ at each end of the resonator. It is given by
\begin{equation}
\label{eqn:spectral-density1}
J_{2}\left(\omega\right) =
\frac{e^2\Omega_2}{\ell c}
\frac{\Gamma_2}{(\omega-\Omega_2)^2+(\Gamma_2/2)^2}
\end{equation}
were $\Omega_2$ is the resonator frequency, $\ell$ is resonator length, $c$ is the capacitance 
per unit length of the transmission line, $C_g$ is the gate
capacitance, and $C_J$ is the junction capacitance. 
The quantity $\Gamma_2 = \Omega_2/Q$ where $Q$ is the quality factor of the cavity. 

For negligeable $Y_{int} (\omega)$, the spectral density of the phase qubit 
shown in Fig.~(\ref{fig:three}) is 
\begin{equation}
\label{eqn:spectral-density-isolation}
J_{3}\left(\omega\right) = \left(\frac{\Phi_0}{2\pi} \right)^2
\frac{\alpha_3\omega}{\left(1-\omega^2/\Omega_3^2\right)^2+4\omega^2\Gamma_3^2/\Omega_3^4},
\end{equation}
where $\alpha_3=L^2/((L+L_1)^2 R) \approx (L/L_1)^2/R$ is the leading order term in the 
low frequency ohmic regime, $\Omega_3=\sqrt{(L+L_1)/(LL_1C)} \approx 1/\sqrt{LC}$ is essentially 
the resonance frequency, and $\Gamma_3=1/(2CR)$ plays the role of resonance width. 
Here, we used $L_1 \gg L$ corresponding to the relevant experimental regime.

Once the spectral functions of the environments are known, the relaxation rates
are derived following standard methods~\cite{weiss-1999}.
At finite temperatures, the relaxation time
\begin{equation}
\frac{\hbar}{T_{1,i}}=\beta_i J_{i}(\omega) \coth \frac{\hbar\omega}{2k_B T}
\label{eqn:relaxation-time}
\end{equation}
is directly related to the environmental spectral density.
Here, the index $i = 1,2,3$ labels the results for flux, charge and phase 
qubits, respectively. The parameter $\beta_i$ takes values $\beta_1 = 1 $ for the flux qubit
of Fig.~\ref{fig:one}, $\beta_2 = C_g^2/(C_g+C_J)^2$ for the charge qubit of Fig.~\ref{fig:two} 
and $\beta_3 = 1/\left[\left(\Phi_0/2\pi\right)^2C_0\omega_{01}\right]$ 
for the phase qubit of Fig.~\ref{fig:three}.  An inspection of Eq.~(\ref{eqn:relaxation-time}) 
and the corresponding spectral functions $J_i (\omega)$ shows that suitably detuning 
the qubit to higher frequencies beyond the environmental resonance can enhance the 
characteristic spontaneous emission lifetime (relaxation time) $T_{1,i}$ of the qubit
by a couple of orders of magnitude as discussed in Ref.~\cite{kaushik-2007}.

\end{document}